\documentclass[a4paper,11pt, twocolumn]{article}
\usepackage{graphicx}
\usepackage{amsfonts}
\usepackage{amsmath}
\usepackage{listings}
\usepackage{mathtools}

\usepackage{hyperref}

\author{Dunn E. J.$^{1,2}$, Young R. J.$^{1}$ and Jarvis S. P.$^{1}$}
\date{%
   \footnotesize $^1$Physics Department, Lancaster University, Lancaster, LA1 4YB, UK.\\%
   $^2$Univ. Lille, CNRS, Centrale Lille, Univ. Polytechnique Hauts-de-France, UMR 8520 - IEMN - Institut d'Electronique de Microélectronique et de Nanotechnologie, F-59000 Lille, France. \\
}
\normalsize
\usepackage[font=small,labelfont=bf]{caption}
\usepackage{xr}

\graphicspath{{Figures/}}

\begin{document}
\title{Single atom chemical identification of TMD defects in ambient conditions}

\twocolumn[
  \begin{@twocolumnfalse}
    \maketitle
    \begin{abstract}
      \noindent The presence of defects in transition metal dichalcogenides (TMDs) can lead to dramatic local changes in their properties which are of interest for a range of technologies including quantum security devices, hydrogen production, and energy storage. It is therefore essential to be able to study these materials in their native environments, including ambient conditions. Here we report single atom resolution imaging of atomic defects in MoS$_{2}$, WSe$_{2}$ and WS$_{2}$ monolayers carried out in ambient conditions using conductive atomic force microscopy (C-AFM). By comparing measurements from a range of TMDs we use C-AFM to chemically identify the most likely atomic species for the defects observed and quantify their prevalence on each material, identifying oxygen chalcogen substitutions and transition metal substitutions as the most likely, and most common, defect types. Moreover, we demonstrate that C-AFM operated in ambient environments can resolve subtle changes in electronic structure with atomic resolution, which we apply to WSe$_{2}$ monolayers doped using a nitrogen plasma, demonstrating the capability of C-AFM to resolve electronic, and chemical-specific, details at the atomic scale. 
    \end{abstract}
  \end{@twocolumnfalse}
]

\paragraph{Introduction} 

The properties of transition metal dichalcogenides (TMDs) make them promising candidates for improving existing technologies and enabling those of the future \cite{mainReview, nonLinearOptics}.  TMDs have a range of interesting properties and applications; some monolayers are direct band-gap semiconductors with excellent optical properties \cite{directBandgap}, they can be highly catalytic as a result of chalcogen defects introducing active sites\cite{HER, Ye1, TMD_HER} and, as layered materials, they can be intercalated, allowing the storage of ions in batteries \cite{batteries}, or stacked to form heterostructures with new properties \cite{twistReview}. The disorder introduced by defects can also be harnessed for the creation of security devices \cite{CaoSecurity}, exploiting the presence of the unique arrangement of defects in individual TMD flakes which is reflected in the photoluminescence response \cite{defectsAffectPL, defectsPlasma}.

Scanning probe techniques offer a means of direct real-space imaging providing insight into the defects present after fabrication and the processes through which they are created and transformed. Recently the ability to resolve these defects with atomic resolution has been demonstrated on TMDs and MXenes using conductive atomic force microscopy (C-AFM) \cite{Sumaiya1, Bampoulis2} and higher eigenmode techniques \cite{Severin1}, complementing the traditional Scanning Tunnelling Microscopy (STM) \cite{Barja1, MoSe2_HER, Addou1, annealForDefects, WSe2_defectDensity, Vansco1, Zhang1_STM, He_Ion_STM, Peto1}.

In this work, the intrinsic defects of mechanically exfoliated TMD monolayers are characterised using C-AFM with atomic resolution in ambient conditions. By comparing results from MoS$_{2}$, WSe$_{2}$ and WS$_{2}$ monolayers, we are able to identify the most likely chemical origins for single atom defects and quantify their prevalence on each material. Moreover, we demonstrate that C-AFM operated in ambient environments can not only resolve defects with atomic resolution, but also detect subtle changes in electronic structure. We apply this method to WSe$_{2}$ monolayers intentionally doped through exposure to nitrogen plasma, where we show that substitutions of individual oxygen and nitrogen atoms on chalcogen sites (O$_{Se}$ and N$_{Se}$) can be distinguished. \\

\paragraph{Methods}
\paragraph{Sample preparation.}

To fabricate TMD monolayer samples bulk TMD crystals (MoS$_{2}$ - Manchester Nanomaterials, WSe$_{2}$ - HQ Graphene, and WS$_{2}$ - 2D Semiconductors) were mechanically exfoliated onto Nitto Blue Tapes (Acrylic/PVC) where repeated exfoliation dispersed and thinned the flakes.  The flakes were then exfoliated from the tapes onto pre-prepared PDMS stamps affixed to glass slides where flakes were then assessed for quality using an optical microscope.  Few-layer flakes were selected for further study with Raman and photoluminescence (PL) in a Horiba labRAM instrument using a Laser Quantum 532 nm laser.  Spectra were used to identify monolayers which were then transferred using the PDMS dry transfer technique onto a freshly exfoliated highly ordered pyrolytic graphite (HOPG) substrate.

Here we use HOPG which exhibits a near ohmic density of states and can be prepared with large atomically flat terraces \cite{HOPG_Ohm}. In particular, good electrical contact with HOPG ensures that gap states within the TMD materials can be easily measured with C-AFM.

Plasma treated samples were prepared in a custom-built glass barrelled plasma reactor. The glass barrel of the reactor was a QVF process pipe (De Dietrich, UK) – dimensions $500$ mm length, $100$ mm diameter, clamped between two custom-made steel end plates which serve as ground electrodes. The system was evacuated using an Edwards RV3 rotary vane pump (base pressure: c. $2 x 10^{-3}$ mbar). The pressure within the system was monitored by a Thermovac TTR 96N SC Pirani gauge and Display One Controller (Leybold UK Ltd) and an in-line liquid nitrogen cold trap was installed to enhance the system base pressure. Plasma was ignited using a RFG-C-100-13 power generator and automatic matching network (Coaxial Power System Ltd, Eastbourne, UK), operated at a frequency of $13.56$ MHz, connected to a $1$ cm thick copper braid (Tranect Ltd, Liverpool, UK) wrapped around the glass barrel three times.

\paragraph{Conductive atomic force microscopy.}

All scanning probe microscopy (SPM) measurements were collected with a Bruker MultiMode 8 equipped with a Nanoscope V controller housed in the Lancaster ultra-low-noise IsoLab facility. The IsoLab facility is comprised of a dedicated building with three above-ground laboratories each contained in their own separate pod. In the basement of each pod sits a 50-ton concrete isolation block with passive isolation providing exceptional low-vibration environments. For the flattening of monolayer flakes and removal of contaminants, the AFM was operated in contact mode using NuNano Scout 70 probes ($f_0$ $\sim 70$ kHz; nominal spring constant 2 Nm$^{-1}$). Samples were typically then checked in PeakForce mode to ensure no damage to flakes. Conductive atomic force microscopy (C-AFM) was carried out using Nanoworld CDT-FMR conductive diamond-coated Si probes ($f_0$ $\sim 105$ kHz; nominal spring constant 6.2 Nm$^{-1}$). C-AFM measurements were collected by operating in contact mode at a deflection setpoint of 5 nm, corresponding to a force of $\sim 31$ nN. A bias voltage was applied to the sample, with the resulting current measured via the conductive probe using a FEMTO DLPCA-200 amplifier.


\paragraph{Results.} 
\paragraph{Flattening of monolayer flakes.}
Atomic resolution imaging with conductive atomic force microscopy (C-AFM) requires atomically flat and electrically conductive substrates, without which features cannot be resolved. 

Following transfer, the TMD monolayer flakes were flattened using the so-called ‘Nano-Squeegee’ technique \cite{Rosenberger2}. This process removes the majority of contamination underneath the TMD flakes introduced from the transfer method, greatly improving the electrical contact and creating atomically flat regions of TMD for high-resolution C-AFM measurements. Flattening is achieved by operating the AFM in contact mode at a higher deflection set point and repeatedly scanning the surface, ensuring that the TMD flakes remain undamaged (see Figure S1).

\begin{figure}[h]
	\centering
	\includegraphics[width=0.5 \textwidth]{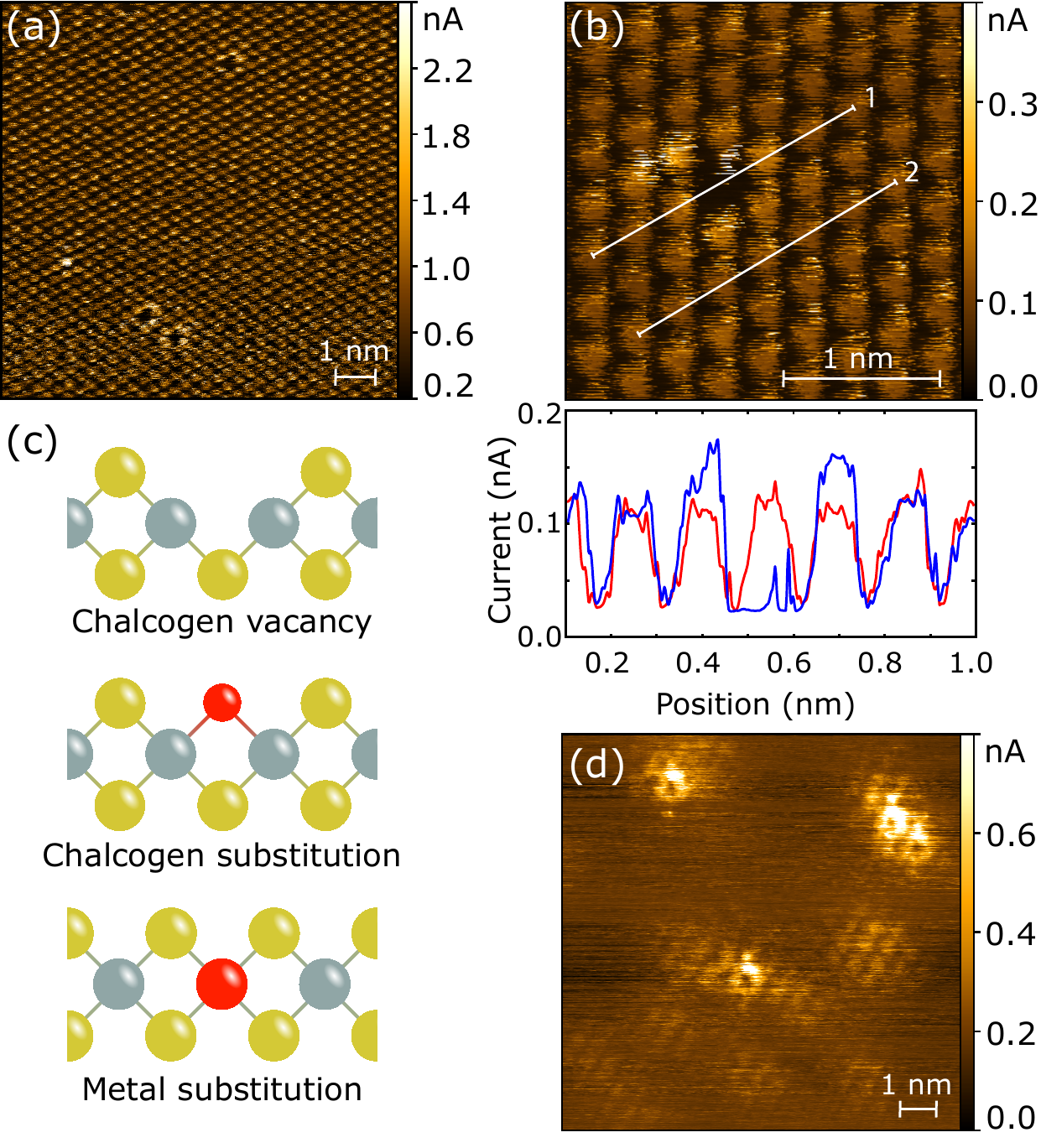} \footnotesize 
		\caption{\label{MoS2} \textbf{Single atom defects in MoS$_{2}$ measured in ambient conditions.} \textbf{(a)} C-AFM image showing atomic scale defects visible as either bright or dark features located on sulfur lattice sites. \textbf{(b)} High resolution images show clear augmentation of the measured current in the nearest neighbour sites around individual defects, with line profile analysis indicating an augmentation of almost $50$\%. \textbf{(c)} Ball-and-stick cartoons depicting the most common defects in MoS$_{2}$. \textbf{(d)} C-AFM image showing `diffuse' defects with large radius.  All C-AFM images in this figure were collected on a MoS$_{2}$ monolayer on HOPG collected at +$0.5\,$V sample bias. }
\end{figure}

\paragraph{Atomic defects in MoS$_{2}$.}
The formation and structure of atomic defects on MoS$_{2}$ is very well studied \cite{Sumaiya1, Addou1, Vansco1, He_Ion_STM, Peto1, Bampoulis1, airStability, oxygenAdsorption, NdopingPlasmaXPS}, and can be used to aid assignment of the defects observed with C-AFM in other materials. In Figure \ref{MoS2} we show the three main defects observed with C-AFM carried out in ambient conditions including images acquired with true atomic resolution.

\normalsize

In Figure \ref{MoS2}a) we show the two most common defects observed, identified as either bright or dark features one atom in size. Furthermore, the surrounding nearest neighbour atoms for these defects appear brighter (increased current) compared to the rest of the MoS$_{2}$ flake. This is most clearly observed for the dark point defects, as shown in more detail in Figure \ref{MoS2}b). The line profile in Figure \ref{MoS2}b) reveals that the current signal is almost 50\% larger for the nearest neighbour atoms compared to atoms further from the defect, suggesting an increase in local unoccupied states due to the presence of the atomic defect, consistent with previous observations on, and simulations of, TMDs \cite{Barja1, Addou1, Vansco1, Peto1, Gonzalez1}. The precise origin of these single-atom size defects is not universally agreed, with reports suggesting that either sulfur vacancies \cite{Addou1, Sumaiya1, Bampoulis2, Bampoulis1} or oxygen substitutions \cite{Barja1, He_Ion_STM, Schuler1, Zheng1} are responsible.  Zhang \textit{et al.} \cite{Zhang_VW} suggest that their most common defects are caused by transition metal vacancies because they observe the defect to lie in-between chalcogen sites, but this is inconsistent with our data shown in Figure \ref{MoS2}. The data from Barja \textit{et al.} is particularly compelling, with the combination of STM, dI/dV, and nc-AFM strongly supporting oxygen substitution. Furthermore, the results from Barja \textit{et al.} are similar to our own assignment that the bright and dark atomic defects arise from sulfur defects in the bottom and top layers of the MoS$_{2}$ monolayer, respectively, with enhancement in the bright defect arising from stronger interaction with the substrate. We also note that whilst our samples are stored under vacuum, ambient measurements necessitate significant exposure to atmosphere and so atomic vacancies are unlikely.  Several studies suggest that oxygen substitutions are energetically favourable \cite{SantoshOxAdsorption, IordanidouOxAdsorption} and that the presence of a chalcogen vacancy encourages the adsorption and dissociation of molecular oxygen; so that in the presence of oxygen, chalcogen vacancies assist in their own transformation into oxygen substitutions \cite{airStability, oxygenAdsorption, IordanidouOxAdsorption}.

\begin{figure}
	\includegraphics[width=0.5\textwidth]{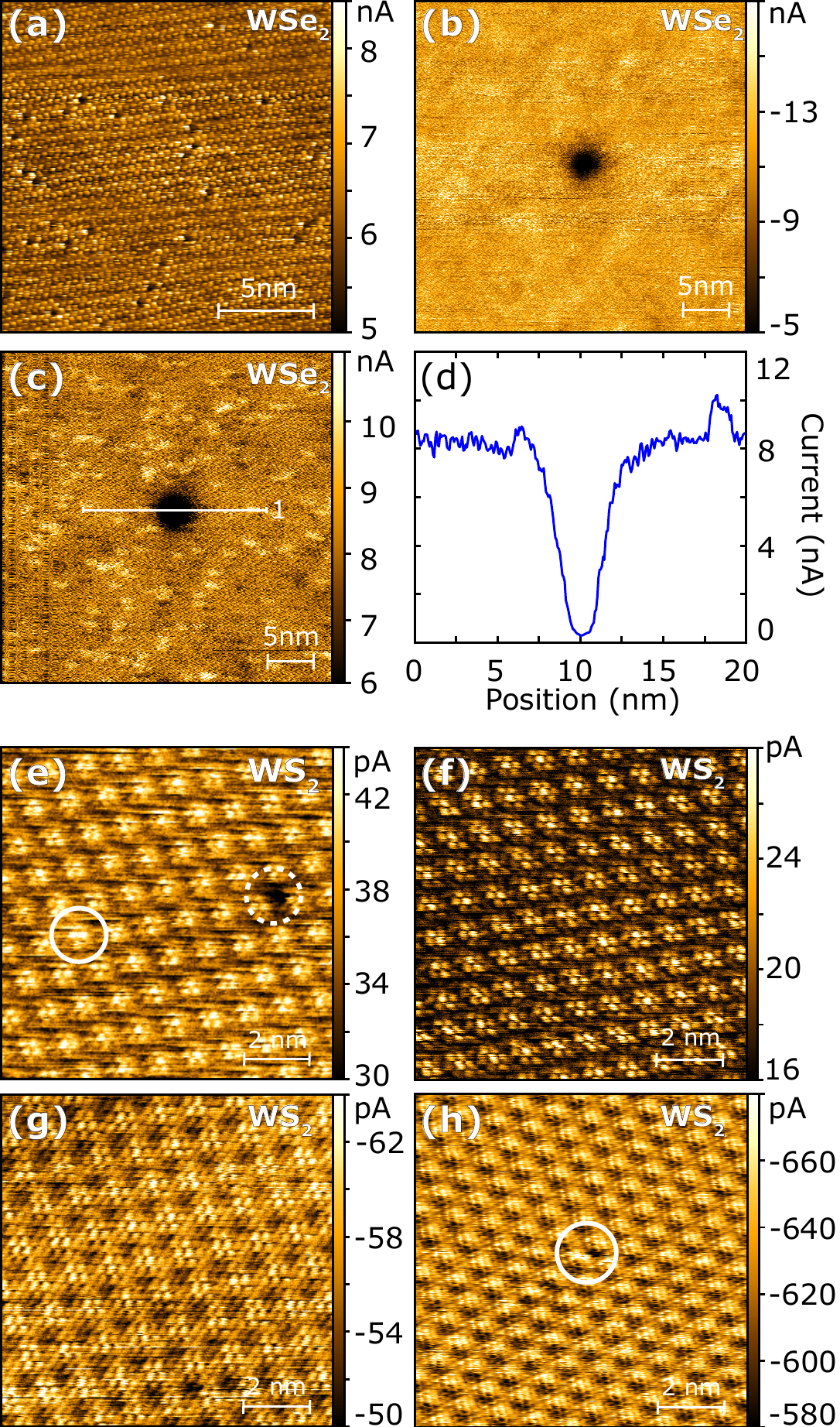} \footnotesize 
		\caption{\label{WSe2} \textbf{Atomic resolution C-AFM measurements of defects in WSe$_{2}$ and WS$_{2}$.} \textbf{(a)} C-AFM image of atomic depleting defects on a WSe$_{2}$ monolayer imaged at $1\,$V. C-AFM image of a diffuse defect on a WSe$_{2}$ monolayer imaged at \textbf{(b)} $-1\,$V and \textbf{(c)} $1\,$V sample bias. \textbf{(d)} A profile across the diffuse defect indicates a diameter of approximately $5$ nm. C-AFM images on WS$_{2}$ are dominated by a Moir\'{e} pattern from the lattice mismatch with the underlying HOPG. \textbf{(e)} Augmenting (white circle) and depleting (dashed white circle) defects on a WS$_{2}$ monolayer observed at $0.5$ V sample bias.  Images collected at \textbf{(f)} $0.5$ V and \textbf{(g)} $-1$ V sample bias in another region show only the Moir\'{e} pattern, with no visible defects. \textbf{(h)} An example of defect, or potentially molecular adsorbate, visible at $-0.8$ V sample bias. }
\end{figure}

Another less commonly observed defect is shown in Figure \ref{MoS2}d). These more diffuse defects appear as either bright (Figure \ref{MoS2}d) or dark (Figure S2) features and are characterised by larger lateral dimensions extending over multiple lattice sites. Diffuse defects have previously been observed in C-AFM and attributed to transition-metal substitutions \cite{Sumaiya1, Bampoulis1}. Similar dark $5$ nm radius defects were imaged via STM by Addou \textit{et al.} who also see significant variation between samples that have been prepared in the same way \cite{Addou1}. Diffuse defects have also been observed in STM by Schuler \textit{et al.} and attributed to charged defects \cite{Schuler_CH}. Charged defects induce band bending, however, in this case the shift is always in the same direction for both the conduction and valence bands. This means that an augmentation of the current at positive (negative) bias leads to a depletion of the current at negative (positive) bias.  Figure \ref{WSe2}b-e) shows that the diffuse defects we observe consistently deplete the current from $-1$V to $1$V, therefore we rule out charged defects as the origin of this effect.

Transition metals can have similar or favourable binding energies when substituted at a molybdenum site when compared to the pristine lattice \cite{Msub}.  Ta, W, Nb, Hf and Zr have higher binding energies than Mo and, as a result, if they are present they could be expected to replace the Mo.  We find that diffuse defects are not observed as commonly on MoS$_{2}$ as the single atom size defects. This is consistent with the expectation that transition-metal substitutions are relatively rare, occurring only during initial growth of the bulk MoS$_{2}$ crystal. These observations therefore support an assignment of diffuse defects to transition metal substitutions, similar to previous reports \cite{Sumaiya1, Bampoulis1}.

\paragraph{Atomic defects in WSe$_2$.}
Beyond MoS$_{2}$, two of the most commonly studied 2D-TMDs are WSe$_{2}$ and WS$_{2}$. There are many STM studies into the chemical nature of defects on WSe$_{2}$ and WS$_{2}$ in synthetic samples \cite{Bampoulis2, Barja1, annealForDefects, WSe2_defectDensity, Zhang1_STM, Schuler1, Schuler_CH, Rosenberger1, Schuler_Bound}. Here we apply C-AFM as described above in order to identify the frequency and type of atomic vacancy observed in mechanically exfoliated monolayers of each material. Figure \ref{WSe2}a) shows a typical atomic resolution C-AFM image collected on WSe$_{2}$ where single atom vacancies are observed. In this case, we observe a much greater number of atomic defects as compared to MoS$_{2}$, most of which appear as dark features which are attributed to Se vacancies or O$_{Se}$ substitutions as described in the previous section. The observation of predominantly dark single atom vacancies is consistent across measurements, implying a greater number of defects in the top layer of the WSe$_{2}$. We tentatively attribute this to increased atmospheric exposure in the top layer, resulting in a larger number of O$_{Se}$ substitutions.

\normalsize

The increased number of single atom features observed on WSe$_{2}$ appears to support reports that defects are more numerous in WSe$_{2}$ due to the reduction in air stability for chalcogens with higher atomic numbers \cite{airStability, airStab_Xnum, TeDegradation}. The comparison between MoS$_{2}$ and WSe$_{2}$ also allows us to provide more insight into the assignment of the atomic defects as oxygen substitutions or chalcogen vacancies. Simulated studies of MoS$_{2}$ and WSe$_{2}$ reveal similar formation energies for chalcogen vacancies in each material. This is due to an increase in binding energy when changing from Mo to W, which acts to counteract the decrease in binding energy when S is exchanged for Se \cite{airStability}. Comparatively, the calculated energy barrier for oxygen adsorption onto WSe$_{2}$ is much less than that required for MoS$_{2}$ \cite{oxygenAdsorption}, suggesting a preference for oxidation on WSe$_{2}$. The increased defect density we observe on WSe$_{2}$ compared to MoS$_{2}$, therefore, suggests that oxidation is most likely responsible, further supporting the assignment of oxygen substitutions rather than Se vacancies.

A higher density of diffuse defects is also observed on WSe$_{2}$. As with other samples, diffuse defects are still much less common than site-specific defects; an observation in agreement with Yankowitz \textit{at al.} \cite{WSe2_defectDensity}. A pair of C-AFM images collected at $\pm 1$ V bias show one such defect in Figure \ref{WSe2}b-c). These defects have a similar lateral size to those on MoS$_{2}$ and, similarly, they may be found as both dark or bright features (Figure S2); though an individual defect is consistently dark or bright across the voltage range used. We note that all C-AFM images presented in this paper use a colour scale that depicts larger current magnitudes as bright, and smaller current magnitudes as dark. It is unclear why some diffuse defects deplete the density of states rather than augment it, though many different transition metals could potentially be substituted into the lattice.

In Figure \ref{WSe2}c) which was collected with a sample bias of $1\,$V we also observe defects with smaller radii. These defects are similar to the single atom size defects observed in other images but appear to be around 1-2 nm in size. We believe this is the result of a slightly blunt tip used in Figure \ref{WSe2}b) and c), as later on in the experiment the resolution improved, allowing us to see these defects.

\paragraph{Atomic defects in WS$_2$.}
Figure \ref{WSe2}(e-h) shows typical C-AFM images collected on WS$_{2}$ monolayers also prepared on HOPG. In this case, strong Moir\'{e} patterns between the WS$_{2}$ and the HOPG substrate dominate C-AFM images. Repeat measurements suggest that Moir\'{e} patterns for WS$_{2}$:HOPG are almost ubiquitous, and are very rarely absent. This can be compared to WSe$_{2}$:HOPG, where Moir\'{e} patterns were also observed, but only at particular bias voltages that could be easily avoided (Figure S3).

Despite complications arising from the Moir\'{e} patterns, defects are still clearly observed, albeit in much smaller numbers compared to the other TMD materials studied. Figure \ref{WSe2}(e-h) shows high resolution C-AFM images collected at a range of sample bias voltages between -1 V and 0.5 V. Small site specific defects are highlighted in Figure \ref{WSe2}(e) showing an augmenting (cyan circle) and depleting defect (red circle) in a C-AFM image collected at 0.5 V sample bias. In this case the Moir\'{e} pattern prevents any possibility for atomic resolution, and the defects instead appear to influence a lateral region of $\sim 1$ nm. Figure \ref{WSe2}(g,h) shows two images from a separate measurement session with only the Moir\'{e} pattern visible collected at 0.5 V and -1.0 V sample bias respectively. A further example of the difficulty in identifying defects can be found in Figure \ref{WSe2}(h) which shows an example of another site specific defect at -0.8 V sample bias. This defect appears to change lattice site during scanning (slow scan direction is top-to-bottom), showing a `slice' effect where the defect switches from a dark to a bright feature, laterally offset by around 0.5 nm. This leads to the conclusion that the defect in Figure \ref{WSe2}(f) may be a surface adsorbate, and so is shown here only for completeness. Diffuse defects, as discussed for WSe$_{2}$ and MoS$_{2}$, were not observed on WS$_{2}$. We tentatively ascribe this to the difficulty in identifying such defects against the background of the Moir\'{e} pattern, and, as discussed below, the relatively low density of defects observed on WS$_2$ compared to the other TMDs studied.

\begin{figure*}[h]
\centering
\includegraphics[width=0.9\textwidth]{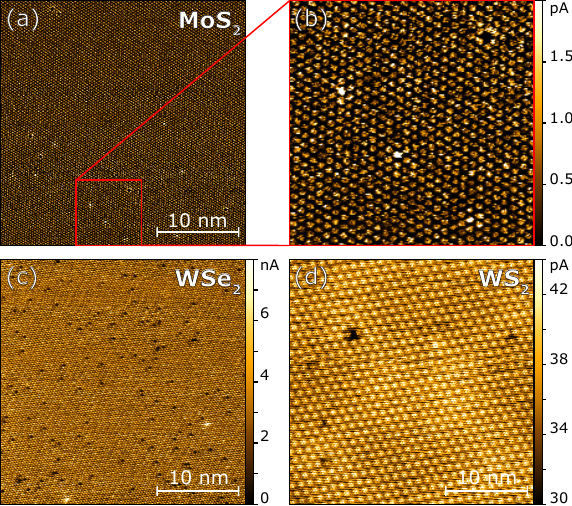}
\caption{\label{densityComparison} \textbf{Quantification of defects on MoS$_{2}$, WSe$_{2}$ and WS$_{2}$.} $30$ nm by $30$ nm current channel C-AFM images of site specific defects on \textbf{(a)} MoS$_{2}$ collected at $0.5$ V sample bias, \textbf{(c)} WSe$_{2}$ collected at $1$ V sample bias, and \textbf{(d)} WS$_{2}$ collected at $0.5$ V sample bias. \textbf{(b)} Shows a section of (a) in greater detail.  The image on WSe$_{2}$ displays the most defects (125), followed by MoS$_{2}$ (16) and then WS$_{2}$ (6), however the Moir\'{e} pattern may hide some of the defects in WS$_{2}$.  In all three materials a mix of augmenting and depleting defects are observed}
\end{figure*}

\paragraph{Quantification of defect densities.}
In Figure \ref{densityComparison} WSe$_{2}$, MoS$_{2}$ and WS$_{2}$ are compared side by side.  Over a $30$ nm by $30$ nm area the defect densities are $1.4 \pm 0.1 \times 10^{13}$, $1.6 \pm 0.4 \times 10^{12}$ and $6 \pm 3 \times 10^{11}$ defects per cm$^{2}$ for WSe$_{2}$, MoS$_{2}$ and WS$_{2}$ respectively. The densities of defects observed in Figure \ref{densityComparison} are consistent with expectations from density functional theory and molecular dynamics studies of oxygen adsorption and dissociation on TMD surfaces and the formation energy of vacancies. Compared to MoS$_{2}$, WS$_{2}$ should exhibit fewer defects due to an increase in the vacancy formation energy arising from the change of Mo to W \cite{airStability}; which then, in turn, decreases the adsorption of oxygen onto the surface \cite{IordanidouOxAdsorption}. Similarly, changing chalcogen from S in WS$_{2}$ to Se in WSe$_{2}$ increases the vacancy formation energy with the same results.  However, it also leads to a higher energy barrier for oxygen adsorption \cite{oxygenAdsorption}.

\paragraph{Nitrogen doping of WSe$_2$ via plasma.}

The atomic resolution capability of C-AFM, and moreover, the sensitivity to minor changes in conduction, suggests that C-AFM should be capable of identifying variations in local conductance by dopant chemistry. WSe$_{2}$ monolayers on HOPG were therefore introduced into a N$_{2}$ plasma, creating N$_{Se}$ substitutions that can be compared with the pre-existing O$_{Se}$ discussed above. XPS measurements have previously confirmed that Mo-N bonds can be introduced to MoS$_{2}$ through plasma exposure, implying that N$_{2}$ plasma creates nitrogen substitutions on the chalcogen sites \cite{NdopingPlasmaXPS, NdopingPlasmaXPS2}, which have a similar formation energy to O$_{Se}$ \cite{NsubFormEnergy}.  

\begin{figure*}[h!]
	\centering
	\includegraphics[width=0.9\textwidth]{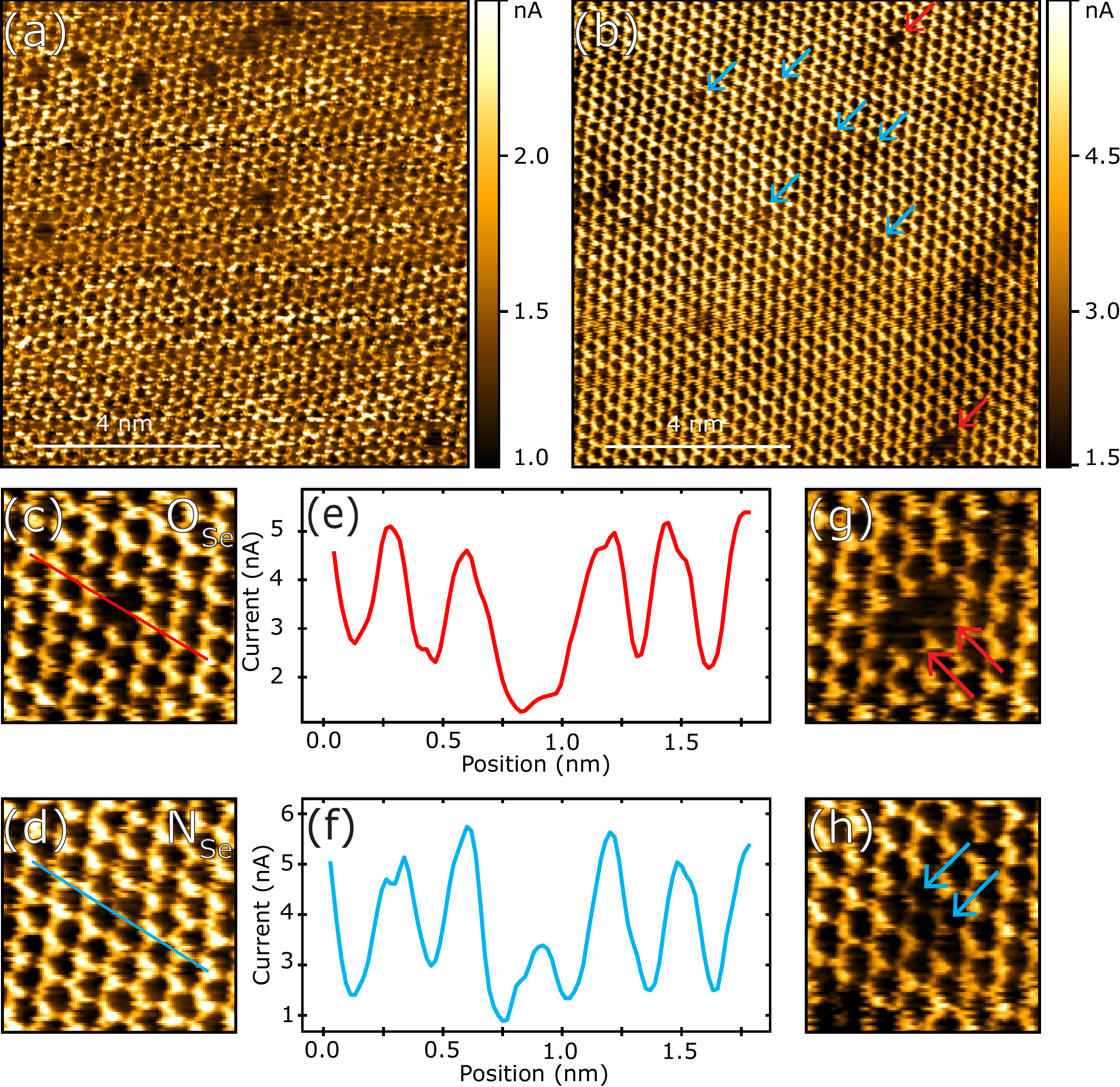}
		\caption{\label{plasma} \textbf{Identification of defects in N$_2$ plasma treated WSe$_{2}$.} Current channel C-AFM images of WSe$_{2}$ monolayer on HOPG imaged at $1$ V \textbf{(a)} before and \textbf{(b)} after plasma exposure. Enlarged C-AFM images show \textbf{(c)} O$_{Se}$ and \textbf{(d)} N$_{Se}$features in detail, alongside line profiles shown in \textbf{(e)} and \textbf{(f)}, respectively. \textbf{(g)} C-AFM image of a pair of O$_{Se}$ defects, compared to \textbf{(h)} a C-AFM image of a pair of N$_{Se}$ defects. Arrows are used throughout the figure to highlight O$_{Se}$ (red) and N$_{Se}$ (blue) features.}
\end{figure*}

Samples were prepared using a bespoke plasma reactor designed for plasma polymerisation experiments which require cleaner vacuum conditions than a typical plasma oven. The number of N$_{Se}$ substitutions can be carefully controlled by varying gas pressure, power, and the duration of plasma treatment. Using conditions of 10 Pa pressure and 10 W power at 13.56 MHz, we found that 10 s of exposure produced a similar density of N$_{Se}$ defects compared to the pre-existing O$_{Se}$. Figure \ref{plasma}a shows a C-AFM image of single atom O$_{Se}$ depleting defects on WSe$_{2}$:HOPG observed before plasma exposure. Following 10 s exposure, a second type of defect appears as shown in Figure \ref{plasma}b, which once again appears as a single atom depleting site, however, compared to O$_{Se}$, a faint atomic feature is still visible. Profiles taken across the defects are shown in Figure \ref{plasma}(e) and (f) alongside enlarged C-AFM images of O$_{Se}$ and N$_{Se}$ in Figure \ref{plasma}(c) and (d), respectively. The line profiles clearly show an increased signal in the depleting region for N$_{Se}$ compared with O$_{Se}$, which we find is consistent across samples (and increase in number with increased plasma exposure). In addition, we note that prior to plasma treatment, defects were not frequently observed at nearest-neighbour chalcogen sites, however, in Figure \ref{plasma}b paired defects are more common and can be observed for both `faint' and single atom depleting defects.  Enlarged examples are shown in Figure \ref{plasma}g-h. 

It has been reported that nitrogen substitutions should add in-gap states near the valence band \cite{NsubFormEnergy}, however, this doesn't appear to be reflected in our C-AFM measurements. We postulate this is because unlike O$_{Se}$, N$_{Se}$ is not iso-electronic, instead acting like a p-type dopant. The underlying HOPG provides access to a carrier reservoir that can fill states, resulting in additional electrons at the dopant site and a local negative charge.  As previously discussed, a key signature of a charged defect is asymmetry between negative and positive bias, therefore, at positive bias a negatively charged defect should be observed as a dark feature, which is consistent with the reduction in current characterising the `faint' defect sites at 1 V. This idea is supported by similarities to charged defects observed in STM studies of CVD TMDs with CH$_{Se}$ and CH$_{S}$ substitutions which are thought to be similar to nature to N$_{Se}$ substitutions\cite{schulerIntentional}.

\paragraph{Mechanism of atomic resolution C-AFM in ambient conditions.}
The diamond-coated probes used in this study last long enough for a full day of measurements, whereas Pt-coated probes could only acquire a small number of images before the conductive signal was lost. Therefore for practical reasons diamond-coated probes were used. Initially, atomic resolution was not expected, however, the diamond probes used achieved `true' atomic resolution in approximately $40$\% of C-AFM imaging sessions, and lattice resolution reliably achieved over $90$\% of the time. We note that the probes remained sufficiently sharp for atomic resolution despite a colossal loading force compared to other high resolution imaging techniques \cite{tipApproach}. We expect that the high stability of the polycrystalline diamond coating plays an important role in maintaining the sharpness of the tip apex. As discussed above, that atomic resolution is not always achieved and despite the high stability of the diamond probes `tip changes' are often observed; generally resulting in a change in image quality and potentially atomic resolution. Tip changes are observed as sudden, discontinuous jumps in the current which originate from changes in the nanostructures at the tip apex or from picking up and dropping molecules on the surface.  In such cases where atomic resolution is not achieved, we find that the majority of defects appear as bright features around 1-2 nm in size, and we no longer observe a mixture of bright and dark point defects. This appears to be consistent with previous reports on CVD WS$_{2}$ using C-AFM with nanoscale resolution \cite{Rosenberger1}.  We also note that whilst attempts were made to measure I(V) at specific defect sites, this data was not reliable due to the high thermal drift in ambient conditions.

High-resolution AFM in ambient conditions has been reported by a number of groups\cite{Sumaiya1, Bampoulis2, Severin1, Bampoulis1, Wastl1, Wastl2, Schimmel1, BetonHighAspect1, BetonHighAspect2, BetonNanowires} with various proposed imaging mechanisms. On TMD samples both Sumaiya \textit{et al.} \cite{Sumaiya1} and Nowakowski \textit{et al.} \cite{Bampoulis2} attribute their atomic resolution images to an insulating layer around the tip which is broken in a sufficiently small area as to create a single atom pathway into the tip.  Nowakowski \textit{et al.} cite studies of the size of the contact area relative to the area of the conducting channel by Enachescu \textit{et al.} \cite{tipModel1} and Celano \textit{et al.} \cite{tipModel2}.  Enachescu \textit{et al.} tentatively propose an insulating layer and Celano \textit{et al.} show that the conductive area of the tip is smaller than the contact area when tunnelling through SiO$_{2}$ at tip loads from $\sim50$-$150$ nN.  We note that all of the tip loads used by the reports where true atomic resolution has been realised all lie below this range: $\sim2$-$2.5$ nN\cite{Bampoulis2}, $\sim0$ nN \cite{Sumaiya1} and $\sim31$ nN (this work). We also note that atomic resolution in UHV-AFM experiments using Si probes have been extensively studied \cite{custanceManipulation, tipApproach, tipFormSim, tipRotation, probeReactivity}, where it has long been known that a single atom at the tip apex determines the tip-sample interaction \cite{giessiblCharge} (and therefore, what one might consider the tip-sample `contact' in C-AFM), with no need for insulating layers. We therefore propose that an insulating layer is not necessary to achieve atomic resolution in ambient, and that the diamond probe likely contains nanocrystallites at the tip-apex, which behave in a similar way to the nano-asperities known to facilitate atomic resolution in UHV-AFM experiments. Indeed, the type of diamond-coated probes used in this study have been observed to terminate in nanoscrystallites in the $10$ nm regime (See Figure S4).

\paragraph{Conclusions.}
We have shown that C-AFM is capable of routinely collecting atomic resolution images of a variety of TMD monolayers prepared and measured in ambient conditions. Not only is C-AFM sensitive to single atom defects, but also subtle changes in the local electronic structure including that of nearest neighbour atoms. MoS$_{2}$ mechanically exfoiliated monolayers on HOPG substrates are used as a well-studied reference where we identify O$_{S}$ and transition metal substituents as the most likely, and most common defect types. Comparison across MoS$_{2}$, WSe$_{2}$, and WS$_{2}$, monolayers reveal consistent variation in defect density of $1.4 \times 10^{13}$, $1.6 \times 10^{12}$ and $6.7 \times 10^{11}$ defects per cm$^{2}$ for WSe$_{2}$, MoS$_{2}$ and WS$_{2}$ respectively, which correspond well to expected formation energies. Finally, we show that minor variations in electronic structure due to O$_{Se}$ and N$_{Se}$ substitutions can be measured in N$_2$ plasma treated WSe$_{2}$:HOPG monolayers further demonstrating the capability of C-AFM to resolve electronic, and indeed chemical, details at the atomic scale. 

By introducing elements into TMDs intentionally via methods such as the plasma exposure demonstrated here it will be possible to find further evidence to illuminate the chemical origin of the defects observed in this work, identify non-intrinsic defects and establish the impact that these defects have on physical properties of TMDs including the optical emission and catalytic behaviour.

\paragraph{Acknowledgements.}

E.J.D. acknowledges support and funding from the GrapheneNOWNANO Centre for Doctoral Training, EP/L01548X/1. R.J.Y. thanks The Royal Society grant number UF160721, Air Force Office of Scientific Research grant number FA9550-19-1-0397 and Engineering and Physical Sciences Research Council, Grant No. EP/K50421X/1. S.P.J. thanks The Royal Society for grant PI70026, and the Engineering and Physical Sciences Research Council (EPSRC) for grants EP/V00767X/1 and EP/X026876/1, Quantum engineering of energy-efficient molecular materials, QMol.  

\bibliographystyle{vancouver}
\bibliography{references}

\end{document}


\title{Supporting Information: Single atom chemical identification of TMD defects in ambient conditions}

\maketitle

\section{Fabrication}

Flakes are exfoliated onto PDMS first for a variety of reasons.  Firstly, exfoliating 2D flakes directly onto a graphite substrate presents challenges, because when the tape is peeled back it is likely to exfoliate the top layer of the substrate.  PDMS is chosen as an exfoliation substrate because it is an insulator that can conform to a surface.

On an insulator, the photoluminescence signal from the flake is greater because carriers in the conduction band cannot move out of the flake and find a non-radiative decay path.  This means that when potential monolayers are checked using optical spectroscopy there is a strong signal.

Exfoliation onto a glass slide was attempted in order to avoid leaving a polymer residue on the sample, but it was too rigid for transferring the flakes to the graphite substrate for measurement.  Due to its rigidity, only a small portion of the slide is able to come into contact with the substrate when they are pressed together.  By exfoliating onto a polymer stamp and then pressing the stamp into the substrate the stamp is able to deform, enabling the flake to be pressed into the substrate.

\begin{figure}[h!]
\centering{
	\includegraphics[width=0.5\textwidth]{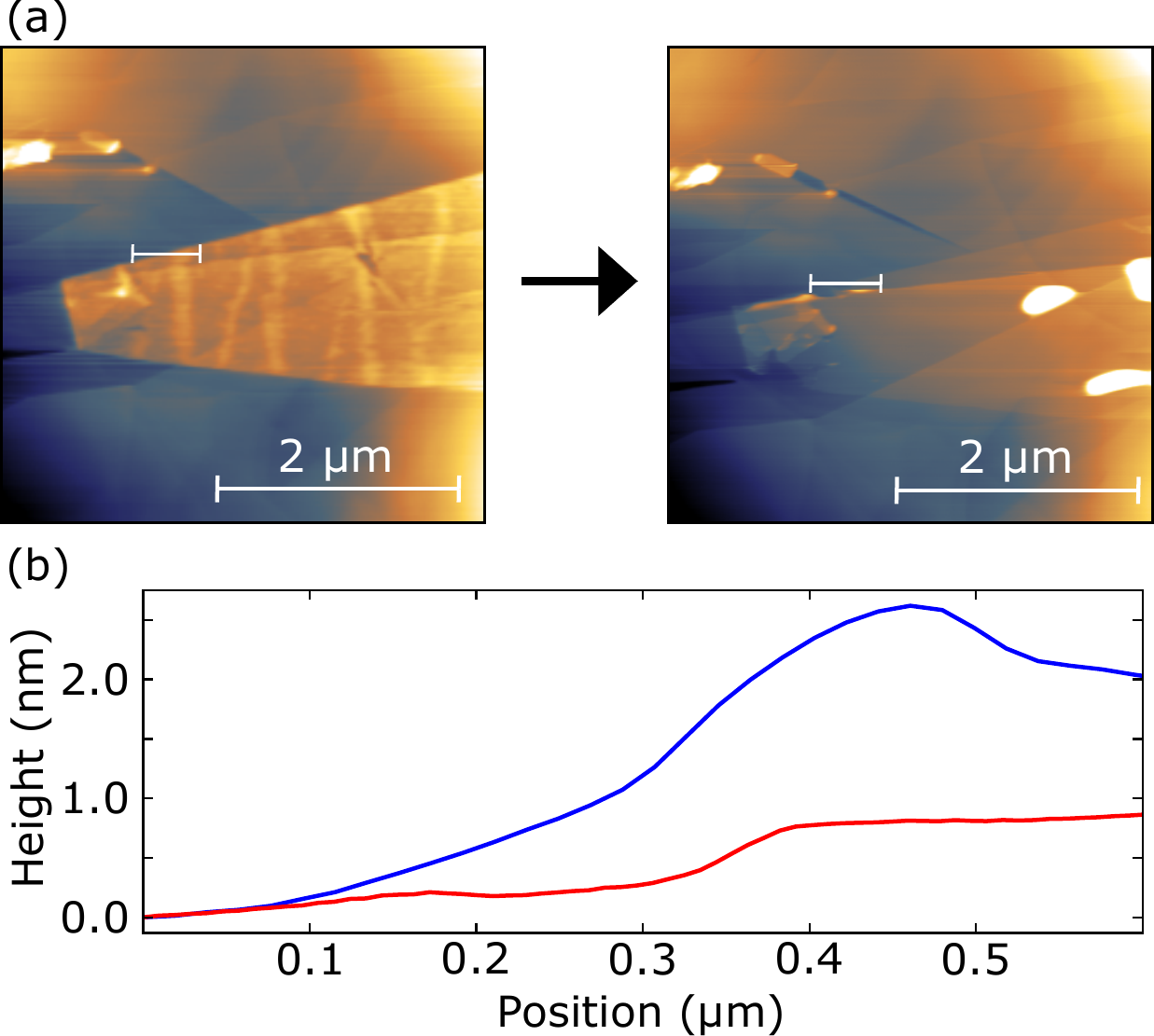}
	\caption{\label{flattening} Contact mode images collected during the flattening of a flake of MoS$_{2}$ \textit{(a)}.  Profiles taken at the same position in both images \textit{(b)} showing that during flattening the apparent height of the monolayer decreases $\sim 2$ nm to just greater than $0.5$ nm, appropriate for a monolayer TMD.}}
\end{figure}

\section{Diffuse Defects}

\begin{figure}[h!]
	\includegraphics[width=\textwidth]{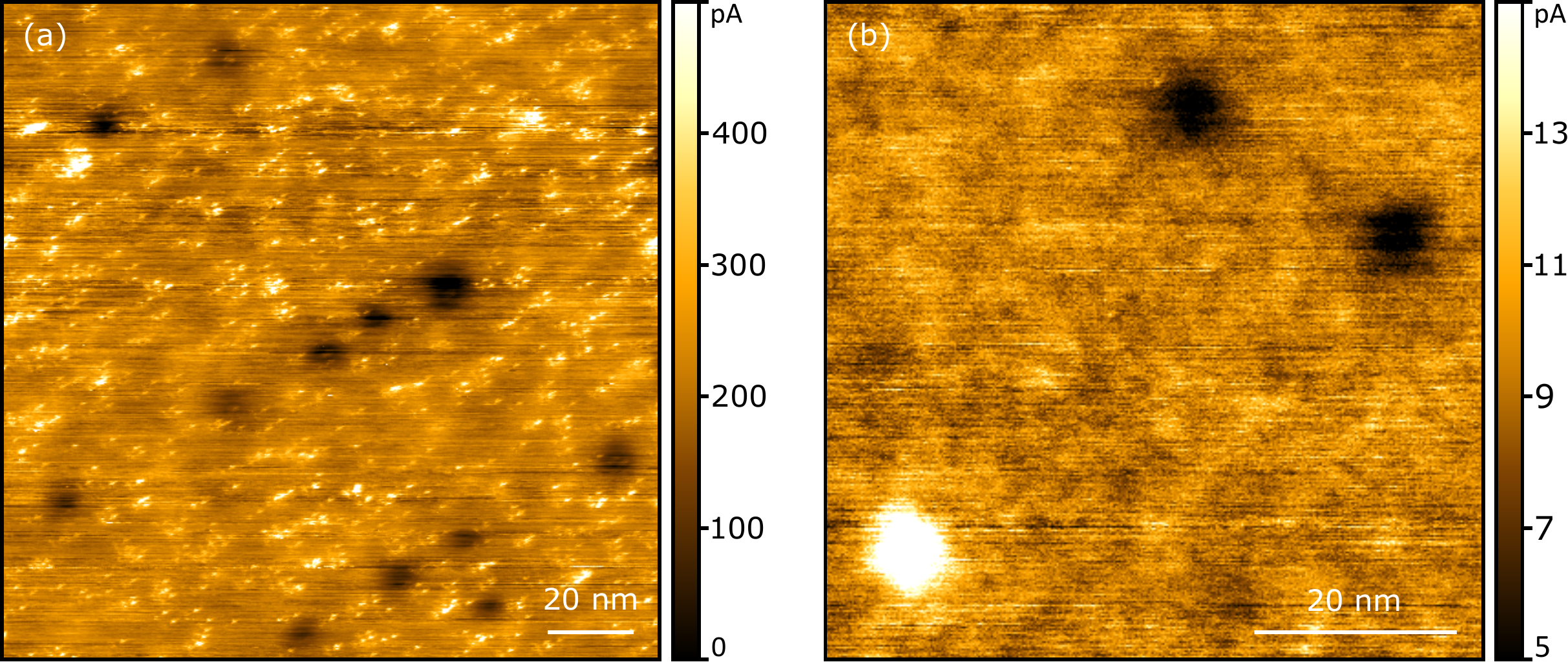}
	\caption{\label{diffuseOpposite} Current channel cAFM images of depleting diffuse defects alongside smaller defects on MoS$_{2}$ \textit{(a)} and alongside an augmenting defect observed on WSe$_{2}$ \textit{(b)}, demonstrating that there are not exclusively augmenting diffuse defects in MoS$_{2}$ or exclusively depleting defects in WSe$_{2}$.  Both images were taken using a bias voltage of $0.5$ V.}
\end{figure}

\noindent Figure \ref{diffuseOpposite} shows diffuse depleting defects on MoS$_{2}$ and both augmenting and depleting defects on WSe$_{2}$.  The diffuse defects in Figure 1 and Figure 2 in the main text are augmenting and depleting respectively.  It is clear that the diffuse defects can be either depleting or augmenting in these crystals.  As diffuse defects are significantly less frequent than single-atom defects, more data would be required to assess whether augmenting or depleting diffuse defects are preferred in a particular TMD crystal.

\newpage

\section{WSe$_{2}$ Moir\'{e} Patterns}

\begin{figure}[h!]
	\includegraphics[width=\textwidth]{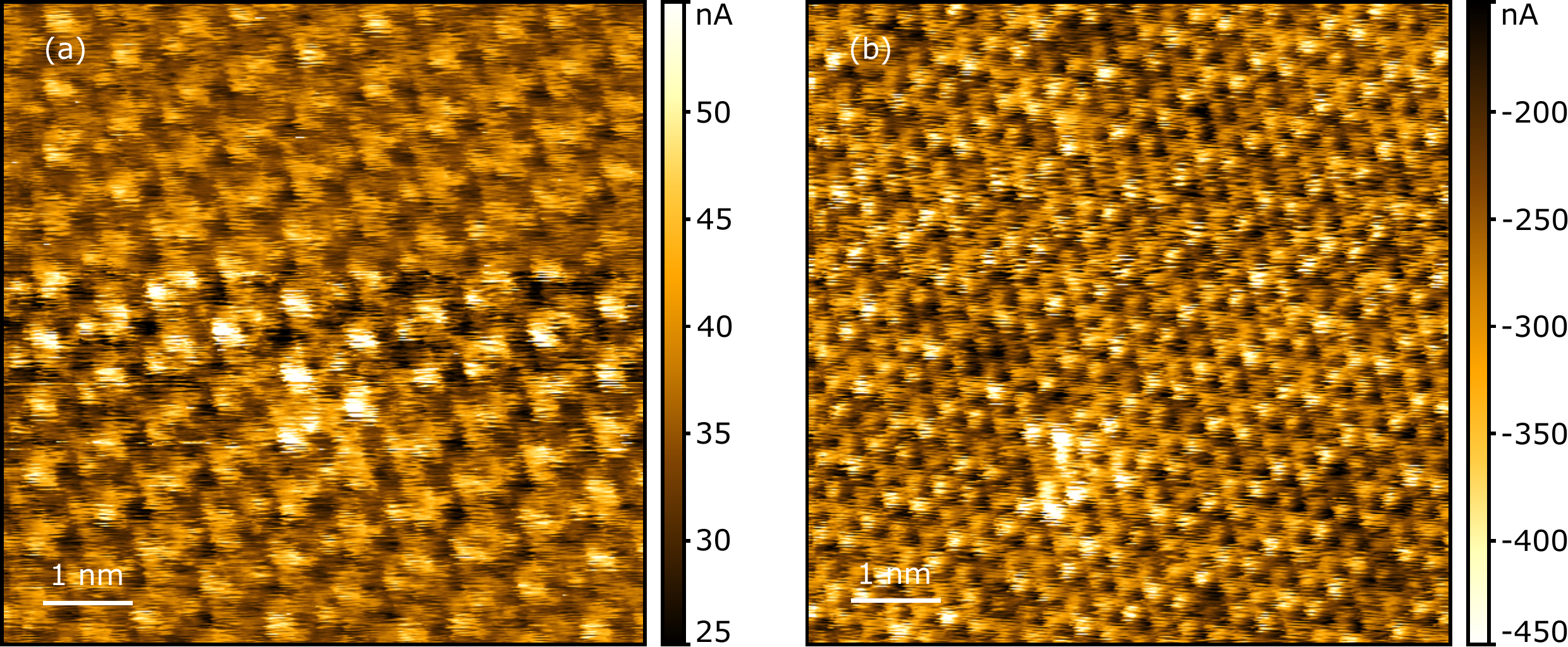}
	\caption{\label{WSe2moire} Current channel cAFM images on WSe$_{2}$.  At a bias voltage of $1$ V \textit{(a)} the periodic pattern is too large to be the lattice and, by changing the bias to $-1$ V \textit{(b)}, the lattice becomes the dominant feature in the current, confirming this.  The image taken at $1$ V bias is likely an image of the Moir\'{e} pattern.}
\end{figure}

Figure \ref{WSe2moire} shows an example of a Moir\'{e} pattern imaged on WSe$_{2}$.  The Moir\'{e} pattern is much less common on WSe2 and MoS$_{2}$ than in images taken on WSe$_{2}$.  The change in the bias voltage from Figure \ref{WSe2moire}a to Figure \ref{WSe2moire}b shows how imaging the Moir\'{e} can be affected by the bias voltage applied.

\newpage

\section{Mechanism of high resolution cAFM imaging}

\begin{figure}[h!]
\centering{
	\includegraphics[width=0.5\textwidth]{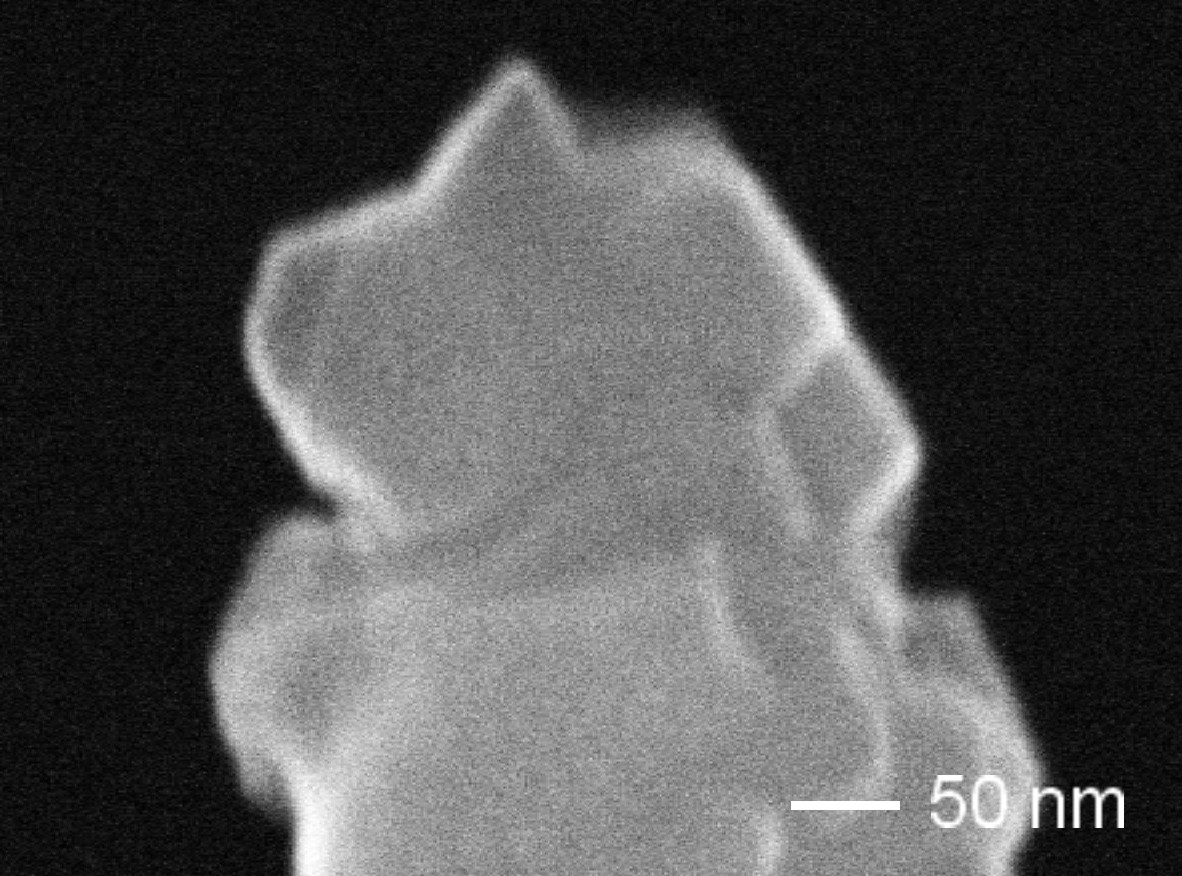}
	\caption{\label{NanoWorldTip} An SEM image showing that nano-crystallites can form very sharp terminations to tips that seem much more rounded.  This image is of a NanoWorld diamond coated AFM tip CDT-FMR. Copyright NanoWorld AG.  This image was provided by NanoWorld and it was included with permission and without modification.}}
\end{figure}

A force of $\sim 31$ nN is very large when applied to the end of a tip.  In order to achieve atomic resolution, it is likely that an atom at the end of the tip needs to protrude slightly beyond the others, dominating the signal due to its marginally closer proximity.  It is reasonable that some proportion of apices will have such a protrusion and high resolution images can be achieved if it is maintained.  If the force is too great the protrusion will be damaged and atomic resolution will be lost.  Diamond is a material with exceptionally strong bonds, but it is possible that our probes only imaged at atomic resolution when the interaction force was anonymously low.

During contact mode AFM the deflection of a probe is taken from the difference between the signal from the top and bottom halves of the photodiode.  As the tip engages this value is used, rather than the difference between the current value and a value when the tip is far from the sample, so the measured deflection is very sensitive to the calibration.  Further to this, the deflection readout can drift significantly while the tip engages.  These factors mean that, though these effects can be minimised, it is difficult to quote an accurate figure for the interaction force in contact mode AFM.

It is important to consider the nature of the measured current when applying a bias voltage in the band gap.  In the traditional view of a semiconductor, the gap is occupied only by defect states.  However, current flows even when the tip is not interacting with a defect and the bias voltage is insufficient for carriers to enter the valence or conduction bands.  It is likely that this current signal originates from the states in the HOPG substrate \cite{MoS2_HOPG}.  In a monolayer TMD, a strong differentiation between the states of the flake and those of the substrate seems unlikely.  It seems reasonable that the appearance of these defects is influenced by some degree of convolution with the HOPG substrate, especially where the signal is as low as it is in the band-gap.

Tip changes are common due to the nature of contact mode AFM as the tip interacts with the surface, picking up contamination or changing the tip shape.  This can cause problems when comparing images, but over a large quantity of images consistent features can be identified and a particular tip artefact will likely change over the course of a day of measurements; revealing the influence of the tip on the images.

\begin{figure}[h!]
	\includegraphics[width=\textwidth]{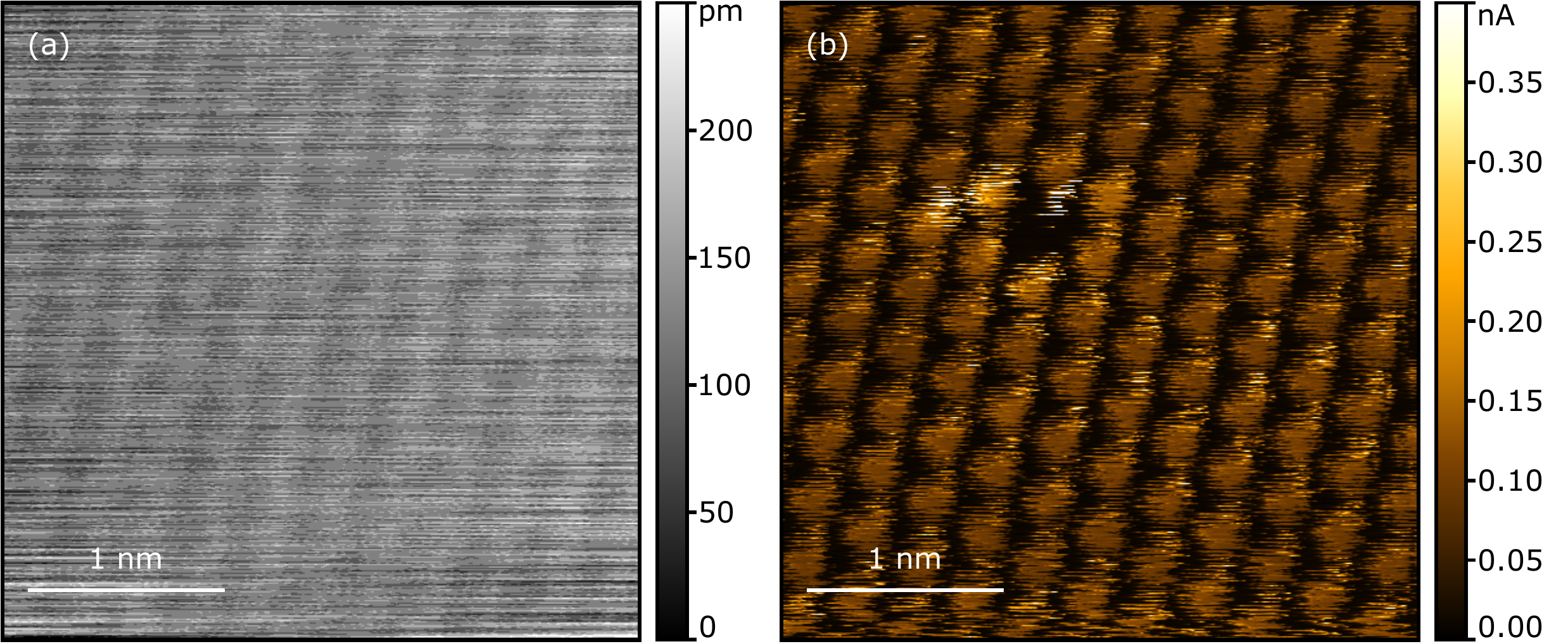}
	\caption{\label{topoVsCurrent} The data from the simultaneously recorded topography \textit{(a)} and current channel \textit{(b)} used to create Figure 1b prior to drift correction.  The lattice can be clearly seen in the topographic image, but the defect is not observed.}
\end{figure}

\section{Nitrogen plasma exposure}

\begin{figure}[h!]
\centering{
	\includegraphics[width=0.5\textwidth]{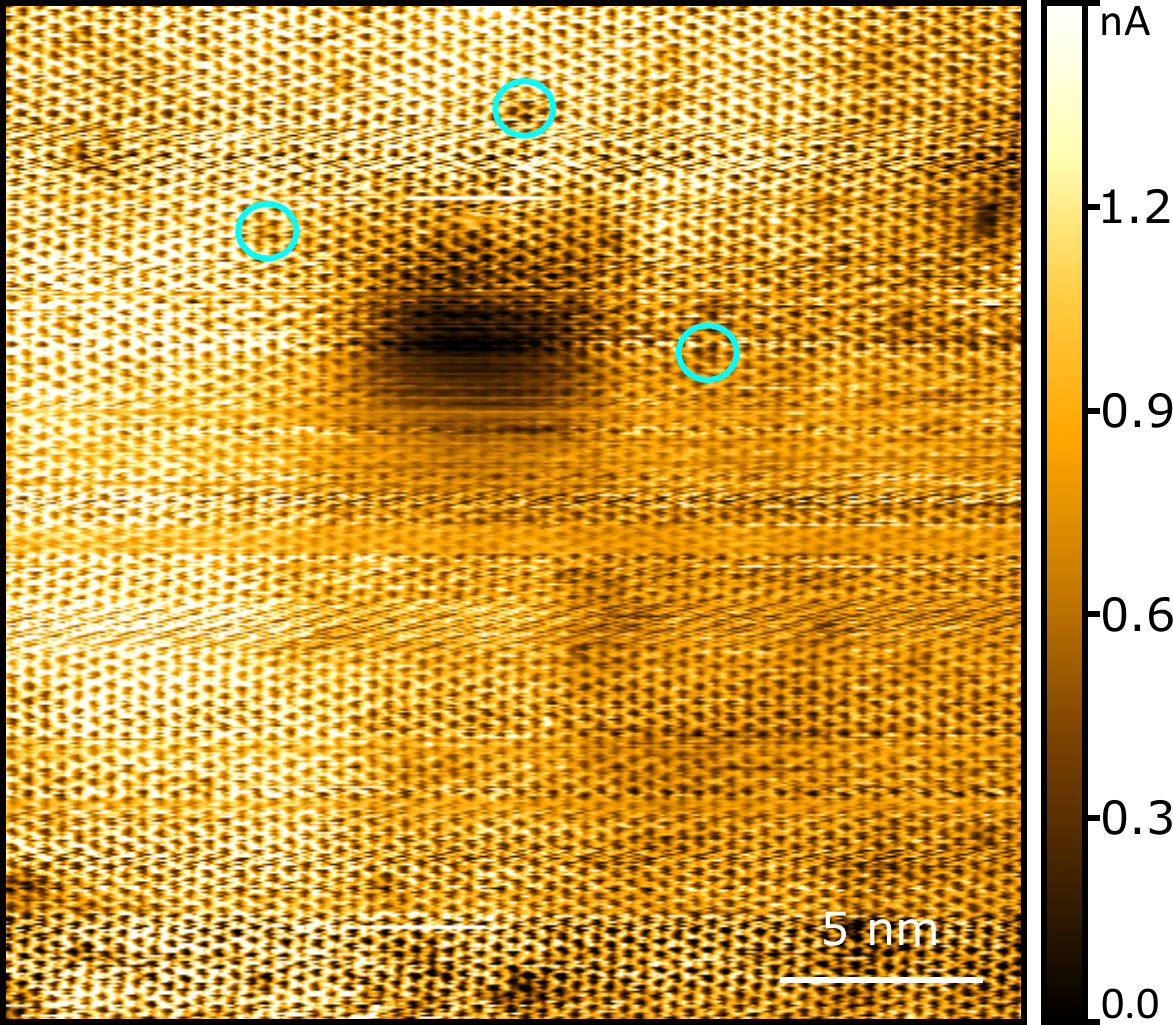}
	\caption{\label{plasmaSI} Atomic resolution current channel image measured at $1$ V bias on monolayer WSe$_{2}$ after plasma exposure.  A diffuse defect can be observed alongside possible N$_{Se}$ defects, identified here by cyan rings.}}
\end{figure}

Despite significant distortion in Figure \ref{plasmaSI}, we can show diffuse defects imaged with atomic resolution.  Alongside the diffuse defects we also resolve single atom sized defects which confirm the resolution.  This is the same sample as is shown in Figure 4.

\section{Uncommon Features}

We do observe some defects in as-exfoliated TMDs which were not reported by Sumaiya \textit{et al.}  In particular, by taking images at a range of bias voltages, we were able to observe defects which are augmenting at large negative bias voltages, but are not visible when this is not the case.  As oxygen substitutions add states at the valence band edge there are clear similarities, but they are not frequent enough for this to be a reasonable assignment.

\begin{figure}[h!]
	\includegraphics[width=\textwidth]{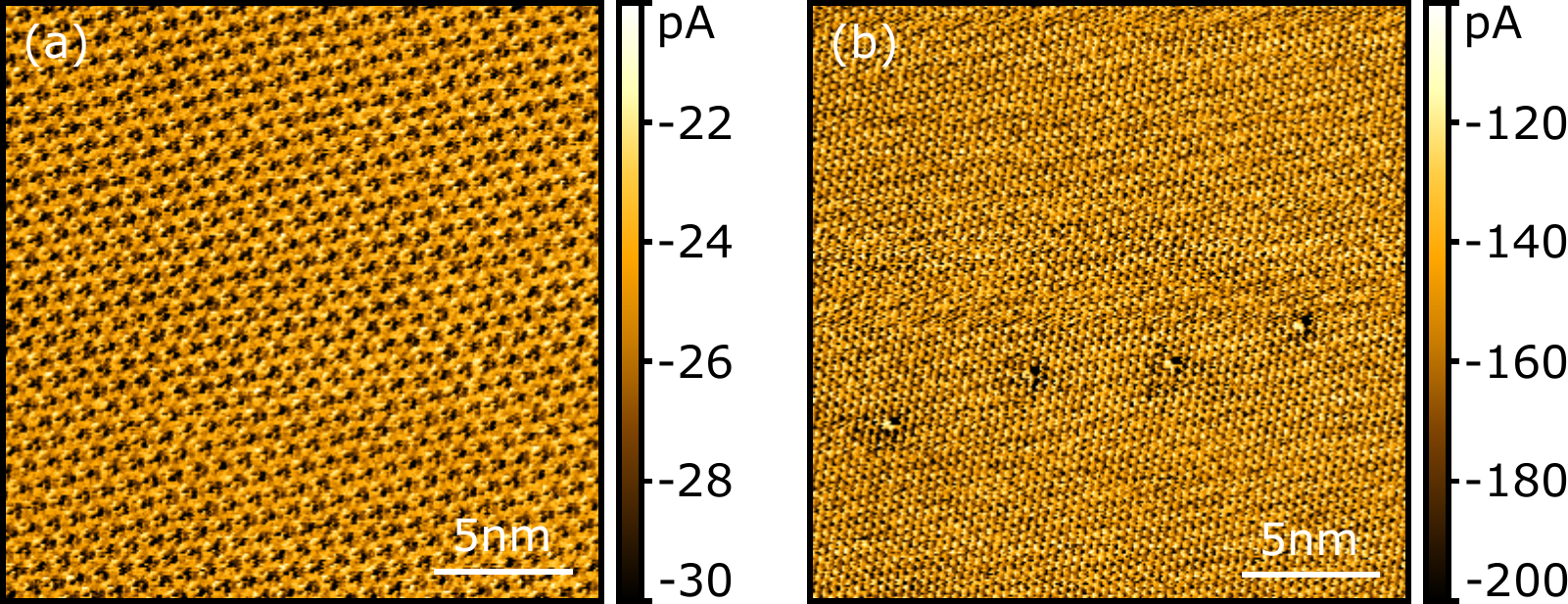}
	\caption{\label{negBiasDefects} A pair of consecutive images of four defects in WS$_{2}$.  Initially, at bias voltage of $-0.4$ V \textit{(a)}, the defects cannot be seen, then when the bias voltage is increased to $-0.8$ V \textit{(b)} they can be made out.}
\end{figure}

\newpage

\bibliographystyle{vancouver}
\bibliography{references}